\documentstyle[prl,aps]{revtex}

\input epsf
\tighten

\begin{document}

\newcommand{\be}{\begin{equation}}
\newcommand{\ee}{\end{equation}}
\newcommand{\ag}{\mbox{I \hspace{-0.82em} H}}
\def\bbox{{\,\lower0.9pt\vbox{\hrule \hbox{\vrule height 0.2 cm
\hskip 0.2 cm
\vrule  height 0.2 cm}\hrule}\,}}

\setlength{\unitlength}{1mm}
\title{Distinguishing between the small ADD and RS black holes in
accelerators }
\author{\\
 Dejan Stojkovic }

\address{ MCTP, Department of Physics, University of Michigan,
      Ann Arbor, MI 48109-1120 USA
}

 \wideabs{
 \maketitle
%%%%%%%%%%%%%%%%%%%%%%%%%%%%%%%%%%%%%%%%%%%%%%%%%%%%%%%
\begin{abstract}
 \widetext
In models with extra dimensions that accommodate a TeV-scale gravity,
 small
black holes could be produced in near future accelerator experiments. Such
small black  holes, whose gravitational radius is mush smaller than the
characteristic size of  extra dimensions (compactification radius in flat
or AdS radius in warped extra dimensions)  can be very
well described by asymptotically flat solutions, thus losing the
 information about the global geometry of the extra manifold. One might
conclude that such  small black holes would be indistinguishable in
different scenarios. We argue that important  differences
still exist, especially regarding experimental signature in colliders,
which may  help us distinguish between the various extra dimensional
 scenarios.
The main differences come from the fact that most of the models with
the warped extra dimension have an additional discrete $Z_2$ symmetry that
makes the brane behave as if it were an infinite tension brane.
\end{abstract}                           }
%%%%%%%%%%%%%%%%%%%%%%%%%%%%%%%%%%%%%%%%%%%%%%%%%%

\narrowtext

Recently, the TeV-scale gravity models have attracted much of
interest.  It is basically  the idea that our $(3+1)$-dimensional
universe is only a sub-manifold on which the standard model fields are
confined inside a higher dimensional space. The original ADD
(Arkani-Hamed, Dimopoulos and Dvali) proposal \cite{ADD} implements
extra space as a multi-dimensional compact manifold, so that our
universe is a direct product of an ordinary $(3+1)$-dimensional
FRW (Freedman, Robertson and Walker) universe and an extra space.
This construction was primarily motivated by attractive particle
physics feature --- namely a solution to the hierarchy problem
(large difference between the Planck scale, $M_{Pl} \sim
10^{16}$TeV and the electroweak scale, $M_{EW} \sim 1$TeV). By
allowing only geometrical degrees of freedom to propagate in
extra dimensions and making the volume of the extra space large,
we can lower a fundamental quantum gravity scale, $M_*$, down to
the electroweak scale ($\sim$ TeV). The size of extra dimensional
manifold is then limited from above only by short distance gravity
experiments (current experiments do not probe any deviations from
a four-dimensional Newton's gravity law on distances smaller than
$0.2$mm). Thus, for different numbers of extra dimensions
compactified on a flat manifold (for an alternative  way of
compactification see \cite{CHM,Dienes}) the compactification
radius can vary from the fundamental length scale $M_*^{-1}$ to
the macroscopic dimensions of order $0.2$mm.

 The other option, exercised in \cite{RSI}, uses a non-factorizable
geometry with a single extra dimension. In the
so-called RSI (Randal, Sundrum) scenario
extra dimension is made compact by introducing two branes (one with
positive and one with negative tension) with a piece of anti-de Sitter
space between them. If we put all the standard model fields on the
negative tension brane, due to exponential scaling properties of masses in
this background, we can solve the hierarchy problem by setting the
distance between the two branes  only one or two orders of magnitude
larger than the anti-de Sitter radius. In order to make a model
selfconsistent, one has to impose a $Z_2$ symmetry around both branes.

Alternatively, we can put all the standard model fields on the
positive tension brane and make the extra dimension infinite by
moving the negative tension brane to infinity (so called RSII
scenario \cite{RSII}). Since this model does not yield a TeV strength
gravity we will confine our discussion (except for the comment at the end)
to the ADD and RSI models.

Large black holes in these two scenarios whose gravitational radius in
brane directions is much larger that the size of extra dimensions
should have properties similar to those of ordinary $(3+1)$-dimensional
black holes. Intermediate size black holes whose gravitational radius is
of the order of the characteristic length of extra dimensions
 (compactification
 radius in ADD or AdS radius in RS  model) can have
quite different
properties due to the ``edge effects" and different
geometry of the extra space. We will not discuss these two
regimes.

Finally, if a gravitational radius of a black hole is much
smaller than the characteristic length of extra dimensions, 
then the black hole can be very
well described by asymptotically flat solutions, i.e. Tangherlini
 \cite{Tang} or
Myers-Perry \cite{MP} solutions for higher dimensional
 static and rotating
 black holes respectively. Thus from this point of view, one might
conclude that there should not be any practical difference between the
small black
holes in these two scenarios. The aim of this paper is to
point out that
important differences, most of them concerning the
experimental signatures in near future accelerator experiments, still
exist. They stem from the fact that RS models usually have a discrete
 $Z_2$
symmetry that fixes the brane and makes it the boundary of space-time.
Such brane behaves as if it were an infinite tension brane for the
processes of interest here.

Probably the most interesting and intriguing feature of  theories with
 TeV-scale gravity is the possibility of production of mini black holes in
future  collider  experiments (for recent reviews see \cite{rev}).
Calculations \cite{Dim} indicate that the probability for
creation of a mini black hole in near future hadron colliders  such as the
LHC (Large Hadron Collider)  is so high that they can  be called  ``black
hole factories".  Consider two particles (partons  in the
case of the LHC) moving in opposite direction with the center of
mass energy $\sqrt{\hat{s}}$. If the impact parameter is less
than  the gravitational radius $r_g$ of a $(N+1)$-dimensional black
hole, then a black hole with a mass of the order of $\sqrt{\hat{s}}$
will form.

For high energy scattering of two
particles with a non-zero impact parameter, the formation of a rotating
black hole is much more probable than the formation of a non-rotating
black hole. One may expect that mainly highly rotating mini black holes
are to be formed in such scattering. To simplify calculations of cross
section of mini black hole production, effects connected with
rotation of the black hole are usually neglected. The same
simplification is usually made when quantum decay of mini black
holes is discussed. However, we argue that rotation is of crucial
interest if we want to distinguish between the small ADD and RS black
holes.

After the black hole is formed it decays by emitting Hawking radiation
 with temperature $T  \sim 1/r_g$ .
Thermal Hawking radiation consists of two parts: (1) particles
propagating along the brane, and (2) bulk radiation. The bulk
radiation includes bulk gravitons. Usually the bulk radiation is
neglected. The reason is as following. The wavelength of emitted
radiation is larger than the size of the black hole, so the black hole
will behave as a point radiator radiating mostly in s-wave.
Thus, the radiation for each particle mode will be equally probable in
every direction (brane or bulk). For each particle that can propagate in
 the bulk there
is a whole tower of bulk Kaluza-Klein excitations, but since they
are only weakly coupled (due to small wave function overlap) to the
small black hole, the whole tower counts only as one particle. Since
the total number of species which are living on the brane is quite large (
$\sim 60$) and there is only one graviton, radiation along the brane
should be dominant (see e.g.
\cite{Myers}). This reasoning works very well if the black hole is not
rotating. Rotation can significantly modify the conclusion.

Indeed, the
number of degrees of freedom of gravitons in the $(N+1)$-dimensional
space-time is ${\cal N}=(N+1)(N-2)/2$. For example, for $N+1= 10$ we
have ${\cal  N}=35$.
One may expect that if a black hole is
non-rotating, emission of particles with non-zero spin (e.g. gravitons) is
suppressed with respect to emission of scalar quanta as it happens in
$(3+1)$-dimensional space-time \cite{Page:76} (see also
Section~10.5 \cite{FrNo} and references therein). However, due to
existence of the ergosphere (region between the infinite redshift surface
and the even horizon), a rotating black hole exhibits an interesting
effect known as super-radiance. Some of the modes of radiation get
amplified taking away the rotational energy of the black hole. The effect
of super-radiance is strongly spin-dependent, and emission of higher spin
particles is strongly favored. For extremely rotating black hole the
emission of gravitons is a dominant effect. For example,
$(3+1)$-dimensional numerical calculations done by Don Page \cite{Page:76}
(see also \cite{FrNo}) show that the probability of emission of a graviton
by an extremely rotating black  hole is about 100 times
higher than the probability of emission of a photon or neutrino. In
\cite{Kerr5D}, it was shown that super-radiance also exist in higher
dimensional space-times. Mini black holes created in the high energy
scattering are expected to have high angular momentum. In the highly
non-linear, time-dependent and violent process of a  black hole creation,
up to $30 \% $ of the initial center of mass energy is lost to
gravitational radiation  (this percentage may be even larger in higher
dimensional scenarios due to  larger number of gravitational degrees of
freedom). Since gravitons are not bound to the  brane, most of them would
be radiated in the bulk giving the black hole a non-zero  bulk component
of the angular momentum. For such a black hole, the bulk radiation
may dominate the radiation along the brane, at least
in the first stages of evaporation \cite{recoil}.

The first signature of bulk graviton emission is virtual energy
non-conservation for an observer located on the brane. Also, as a
result of the emission of the graviton into the bulk space, the black hole
recoil can move the black hole out of the brane.
After  the black hole leaves the brane, it cannot emit brane-confined
particles anymore. Black hole radiation would be abruptly terminated for
an observer located on the brane. Probability for something like this to
happen depends on many factors (mass of the black hole, brane tension...)
and it was studied in \cite{recoil}.

Another important question, often neglected in discussion, is the
interaction between the black hole and the brane.
In \cite{friction} the rate of the loss of the
angular momentum of the black hole which interacts
with a stationary brane was calculated. It was shown that a black hole
in its final stationary state can have
only those components of the angular momenta which are connected with
Killing vectors generating transformations preserving a position of
the brane (see Fig. 1.). This is a direct consequence
of the ``friction" between the black
hole and the brane. As a result of this friction the black hole
loses all of the bulk components of its angular momentum to the brane.
The only components of angular momentum which survive are those along the
 brane. The characteristic time when a rotating black hole with
the gravitational radius  $r_g$ reaches this final state
is

\be
(\Delta t)_F \sim r_g^{k-1}/(G_*\sigma) \, ,
\ee
where $G_*$ is the higher dimensional
gravitational coupling constant, $\sigma$ is the brane tension, and
$k$ is the number of extra dimensions.
The rotating black hole can also lose its bulk components of the rotation
by emitting  Hawking quanta in the bulk.  The characteristic time of this
process is $(\Delta t)_{H}\sim t_* (r_g/L_*)^{3+k}$, where $t_*$ and $L_*$
are the fundamental time and length. For black holes which can be
treated classically  $r_g \gg L_*$, so that we have $(\Delta t)_{H}\gg
(\Delta t)_F $.  Thus the friction
effect induced by the brane is the dominant one.

Finally, let us see how previous discussion can help us distinguish
 between the small
ADD and RS black holes in accelerator experiments.
A characteristic property of RS models is the existence of a discrete
 $Z_2$ symmetry with respect to $y \rightarrow -y$ (where $y$ is the extra
 dimension). Under this $Z_2$ transformation the brane remains unchanged,
while the components of any vector orthogonal to the brane change
their sign.  Thus $Z_2$ symmetry implies that any bulk components
of the angular momenta of the small black hole attached on
the brane are strictly zero (see Fig. 1.).  Hence
a stationary black hole attached to the brane in the RS-model can
rotate only within the brane. Exact solutions describing rotating
black holes on ($2+1$)-dimensional branes \cite{EHM} possess this
property. This implies that for a small RS black hole emission of bulk
gravitons would be heavily suppressed since they carry a
non-zero bulk angular momentum. This is in strong
contrast with a small ADD black hole.

\vspace{-1cm}

\begin{figure}[h!]
\label{KillingRot}
\center
\epsfxsize = 0.9\hsize \epsfbox{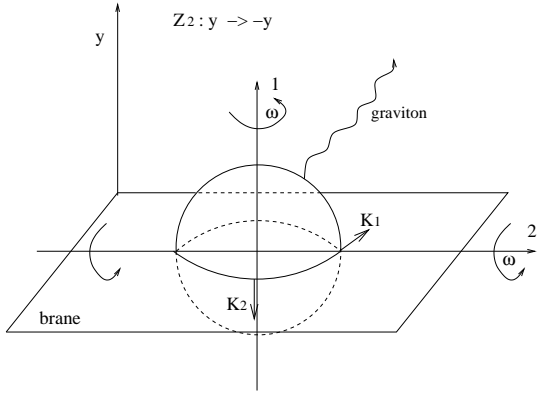}
\vspace{-1cm}
\caption{Due to the ``friction" between the black
hole and the brane, an ADD black hole losses bulk components of the
 angular momentum.
Due to $Z_2$ symmetry an RS black hole can not have any bulk components
of the angular momentum. $K_1$ and $K_2$ are the Killing vectors
generating rotations around axis $1$ and $2$ respectively.}
 \end{figure}

It is not difficult to show formally the absence of bulk rotation for a
small RS black hole on the brane. Consider a small black hole in RS
scenario that can be described by a Myers-Perry solution with two
rotational parameters. The metric is \cite{MP}:

\begin{eqnarray} \label{K5D} &&ds^2=- dt^2 + {r^2\rho^2\over \Delta}\,
dr^2+\rho^2\, d\theta^2\,  +(r^2+a^2)\, \sin^2\theta\, d\phi^2
 \\ &&+(r^2+b^2)\, \cos^2\theta\, d\psi^2 + {r_0^2\over \rho^2}
\left[dt+a\, \sin^2\theta\, d\phi +b\, \cos^2\theta\, d\psi  \right]^2\, .
\nonumber \end{eqnarray}

Here,
\begin{eqnarray}
&& \rho^2=r^2+a^2\,\cos^2\theta+b^2\,\sin^2\theta\, , \\ &&
\Delta=(r^2+a^2)(r^2+b^2)-r_0^2\, r^2\, . \nonumber
\end{eqnarray}

 Angles $\phi$ and $\psi$
take values from the interval $\left[0,2\pi \right]$, while angle
$\theta$ takes values from $\left[0,\pi/2 \right]$.
We specify the position of the brane in the equatorial plane of the black
hole at $\psi =0$. Then, $a$ is the rotational parameter describing
rotations within the brane, while $b$ is the bulk rotational parameter.

The general form of a 5-dimensional metric is:

\be ds^2 = g_{MN} dx^M dx^N \ee
where indices  $M$ and $N$ go over all $5$ dimensions.
We can decompose it into:
\be ds^2 = g_{\mu \nu} dx^\mu dx^\nu + 2 g_{\mu y} dx^\mu dy + g_{yy} dy^2
\, , \ee
where $\mu = 0,1,2,3$, while $y$ is the extra coordinate.

We can now impose a $Z_2$ symmetry under the transformation $y \rightarrow
-y$.  Then, the metric components must satisfy this:
 \be
 g_{\mu \nu} \rightarrow g_{\mu \nu} \, , \ \ \ \
 g_{\mu y}  \rightarrow  - g_{\mu y}  \ \ \ {\rm and} \ \ \
 g_{yy}  \rightarrow g_{yy} \, .
\ee

Imposing a symmetry under $y \rightarrow
-y$ is equivalent to restricting the interval for $\psi$ to
$\left[0,\pi \right]$ and requiring a symmetry under $\psi \rightarrow
-\psi $. From the explicit form of the metric (\ref{K5D}) we see that
in this case the following must be true:

\be a  \rightarrow a \, , \ \ \ \  b \rightarrow -b \, . \ee

If the RS black hole is in the bulk, this would imply that the
($Z_2$-symmetric) ``mirror" image black hole on the other side of the
brane must be spinning in the opposite direction as far as the bulk
angular momentum is concerned. However, if the black hole is located on
the brane, the $Z_2$ transformed metric must describe the same object as
the original one, and we can conclude that $b$ must be zero.

Note that this argument is valid only for a small black hole. If the
extent of the black hole horizon into the extra dimension is larger than
the AdS radius, the symmetry group that describes rotations in $4$ spatial
dimensions is not $SO(4)$ anymore since the fourth dimension is not
equivalent to the first three dimensions. It is still possible that the
same conclusion remains but it requires more careful arguments.

If an ADD black hole emits any particle with a bulk angular momentum, it
would acquire a  bulk angular momentum itself. The same would happen if a
black hole collides with a particle from the bulk. But we saw that RS
black hole can not have any bulk component of angular momentum. This
implies that a $Z_2$ symmetric brane behaves as an infinite tension brane
absorbing all the incoming bulk angular momentum.

As a consequence of a virtual absence of the bulk radiation, a small RS
black hole that is attached to the brane can not recoil and leave the
brane. This is not (quite) surprising since the $Z_2$ symmetry
anyway prevents the black hole from leaving the brane. The process in
which a small RS black hole leaves the brane in the  $Z_2$ symmetric space
where the two sides of the bulk space are not identified reminds of a
black hole splitting into two symmetric black holes  (see Fig.
2.). Classically this process is forbidden in a higher dimensional
space-time for the same reason as in $(3+1)$-dimensional space-time in
connection with non-decreasing property of the entropy. The entropy of a
black hole in fundamental units is  $S \sim  M_{BH}^{\frac{2+k}{1+k}}$,
and the entropy of the final state (two black
holes of mass $M_{BH}/2$) is lower than  that of the initial one.
If the two sides of the bulk space are identified, the ``image" black
hole  is identified with the original one and we do not count degrees of
freedom of two black holes separately. However, even in this case the
recoil effect is suppressed due to highly suppressed bulk radiation.
This is again in strong contrast with a small  ADD black hole.

One may argue that the process of the extraction of the black hole
from the brane is time dependent and that there will be some energy flux
through the horizon during the process. The black hole would grow in mass
 and thus we could go around the argument of the smaller entropy in the
final  state. In \cite{flux}, energy fluxes in time-dependent
configurations of a black  hole-brane system were calculated. It was shown
that if the process of extraction is  adiabatic (quasi static)
the energy flux through the horizon is negligible. Thus, in general, we
can not use counter-arguments of this type.

Therefore, even if a black hole emits a particle in the
bulk direction (or gets hit by a particle from the bulk), it will not
recoil off the brane.  Again, it looks as if a $Z_2$ symmetric brane
behaves as an infinite tension brane absorbing all the incoming bulk
linear momentum.

We note here that there is no paradox in these statements.
In  a $Z_2$ symmetric space, it is implicitly assumed that two identical
particles are emitted (absorbed) by a black hole in a $Z_2$ symmetric way,
thus canceling out any bulk components of momentum. In a representation
where only one half of the space is shown (the two sides are identified)
and the brane is a boundary of the space, it looks like the bulk component
of (linear or angular) momentum is absorbed by the brane since there is no
``other side" of the space. Also, a brane in a $Z_2$ symmetric space is
fixed and can not vibrate, thus behaving as if it were a infinite tension
brane though its physical tension can be finite.

\vspace{-4cm}

\begin{figure}[h!]
\label{RSrecoil}
\hspace{-1cm}
\center
 \epsfxsize = 1.10 \hsize \epsfbox{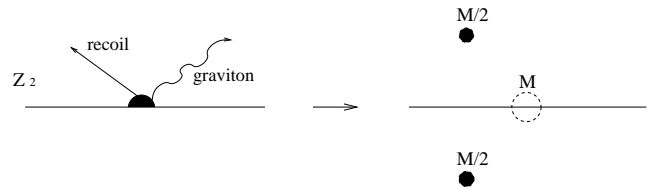}
\vspace{-3cm}
\caption{Recoil of a small RS black hole looks like a black hole
splitting into two symmetric black holes in the ``mirror'' space. It is
not possible to achieve this spontaneously without violating either the
conservation of  energy or non-decreasing property of entropy.}
\end{figure}

We would like to conclude with outlining the main features
that can help us distinguish between the ADD and RS black holes
in near future accelerator experiments.

\underline{A small ADD black hole:}

1. The first phase of Hawking radiation is mostly in the bulk
(the second phase is mostly on the brane)

2. Existence of relaxation time during which the black hole
looses the bulk components of angular momentum

3. A black hole can recoil and leave the brane

\underline{A small RS black hole:}

1. Bulk radiation is strongly suppressed

2. Absence of any bulk components of angular momentum (absence of
 relaxation time)

3. A black hole {\em cannot} recoil and leave the brane

Note that other differences can still exist. For example,
different  Kaluza-Klein spectrum in ADD and RS scenarios can imply
different radiation patterns and/or lifetimes for small black holes
\cite{cas}. This discussion is out of the scope of this paper.

We should add that all of the properties listed above for the small
black holes attached to the brane in RSI scenario apply equally to the
small black holes in RSII scenario (with a $Z_2$ symmetry), except that
they can not be produced in near future accelerator experiments. Instead,
the cases of interest would be the final stages of evaporation of a large
RSII black hole where horizon  shrinks to the size much smaller that the
AdS radius, and small primordial black  holes formed
in energetic processes in the early universe.

Practically all of the facts that make the RS black holes different are
closely connected with a $Z_2$ symmetry. In RSI, we need this symmetry to
fix the brane and prevent dangerous oscillations of a negative tension
brane. In RSII, this symmetry is not necessary, although it simplifies
the model considerably (see for example \cite{acd}). However, if we want
an AdS/CFT interpretation \cite{adscft} of the RSII model, we need a $Z_2$
symmetry to make a brane a boundary of space-time. In any case, a
$Z_2$-symmetric brane behaves as if it were an infinite tension brane for
the black hole processes of interest here.
Keeping an open mind, we should allow for the  possibility that
this symmetry is artificially imposed or maybe a too strong
requirement. If the $Z_2$ symmetry can be relaxed, some of the conclusions in this paper would  change.
If however, the $Z_2$ symmetry is necessary for the self-consistency of
the model than distinct experimental signature in colliders may help us
distinguish between the different extra dimensional scenarios.
In particular they can tell us if we live "on the edge" of the space-time
or not.

\vspace{12pt} {\bf Acknowledgments}:\ \
The author is grateful to Valeri Frolov with whom most of the related
work had been done and Glenn Starkman, Jim Liu and  Dmitri Fursaev for
extremely useful conversations. The work was supported by the DOE grant to
the Michigan Center for Theoretical Physics, University of Michigan, Ann
Arbor.


\begin{thebibliography}{9}



\bibitem{ADD}
N. Arkani-Hamed, S. Dimopoulos and G. Dvali, Phys. Lett. {\bf B429},
263 (1998); I. Antoniadis, N. Arkani-Hamed,  S. Dimopoulos and G. Dvali,
Phys. Lett. {\bf B436}, 257 (1998).



\bibitem{CHM}  N. Kaloper, J. March-Russel, G. Starkman and M. Trodden,
  Phys. Rev. Lett. {\bf 85}, 928 (2000);  G. Starkman, D. Stojkovic and
  M. Trodden, ; Phys. Rev. Lett. {\bf 87}
 231303 (2001); Phys. Rev.
{\bf D63}, 103511 (2001)

\bibitem{Dienes} K. R. Dienes, Phys. Rev. Lett. {\bf 88} 011601 {2002}

\bibitem{RSI}
L. Randall and R. Sundrum, Phys. Rev. Lett. {\bf 83}, 3370
(1999);
\bibitem{RSII}L. Randall and R. Sundrum, Phys. Rev. Lett. {\bf 83}, 4690
(1999);


\bibitem{Tang} F. R. Tangherlini, Nuovo Cim. {\bf 77}, 636 (1963).

\bibitem{MP} R. C. Myers and M. J. Perry, Ann. Phys. (N.Y.) {\bf 172},
304 (1986).


\bibitem{rev} P. Kanti, {\t hep-ph/0402168};  K. Cheung, {\it
hep-ph/0409028}, O.  Dias  {\it hep-th/0410294}


\bibitem{Dim} T. Banks, W. Fischler, {\it hep-th/9906038 };
S. Dimopoulos, G. Landsberg, Phys. Rev. Lett. {\bf 87} 161602 (2001)
;
S. B. Giddings and S. Thomas, Phys. Rev. {\bf D65} 056010 (2002)



\bibitem{Myers} R. Emparan, G. Horowitz, R. C. Myers, Phys. Rev. Lett.
{\bf 85} 499 (2000)


\bibitem{Page:76} D. N. Page, Phys. Rev. {\bf D13} 198 (1976);
    Phys. Rev. {\bf D14} 3260 (1976).

\bibitem{FrNo} V. Frolov and I. Novikov. {\em Black Hole Physics: Basic
Concepts and New Developments} (Kluwer Academic Publ.), 1998.




\bibitem{Kerr5D}
V. P. Frolov, D. Stojkovic, Phys. Rev. {\bf D67} 084004 (2003);
Phys. Rev. {\bf D68} 064011 (2003)


\bibitem{recoil}
V. Frolov, D. Stojkovic, Phys. Rev. Lett. {\bf 89} 151302
(2002); Phys. Rev. {\bf D66} 084002 (2002); M. Cavaglia, Phys. Lett. {\bf
B569} 7 (2003)

\bibitem{friction} V. P. Frolov, D. V. Fursaev, D.
Stojkovic, Class. Quant. Grav. {\bf 21} 3483 (2004); JHEP
{\bf 0406} 057 (2004); V. Frolov, M. Snajdr and D. Stojkovic,
Phys. Rev. {\bf D68} 044002 (2003);

 \bibitem{EHM} R. Emparan, G. Horowitz, R. C. Myers,  JHEP 0001:007
 (2000);      JHEP 0001:021 (2000)


\bibitem{flux} D. Stojkovic, JHEP 0409:061 (2004)

\bibitem{cas} R. Casadio, B. Harms, Int. J. Mod. Phys. {\bf A17} 4635
(2002); P. Kanti, J. Grain, A. Barrau, {\it  hep-th/0501148 };

\bibitem{acd} D. Jennings, I. R. Vernon, A. C. Davis, C. van de Bruck {\it
hep-th/0412281 }


\bibitem{adscft} M. J. Duff, James T. Liu,
Phys. Rev. Lett. {\bf 85} 2052 (2000);
R. Emparan, A. Fabbri, N. Kaloper,  JHEP {\bf 0208} 043 (2002); N.
Arkani-Hamed, M. Porrati, L. Randall,  JHEP {\bf 0108} 017 (2001) ;
 A. Hebecker, J. March-Russell,  Nucl. Phys. {\bf B608} 375 (2001)

\end{thebibliography}
\end{document}